\documentstyle[prl,aps]{revtex}

\begin{document} 
\draft

\title{Inflation in Kaluza-Klein Theory: Relation between
the Fine-Structure Constant and the Cosmological Constant}
\author{Li-Xin Li and J. Richard Gott, III}
\address{Department of
Astrophysical Sciences, Princeton University, Princeton, NJ 08544}
\date{April 28, 1998}
\maketitle

\begin{abstract}
In this paper we investigate a model of an inflationary
universe in Kaluza-Klein
theory, which is a four-dimensional de~Sitter space plus a
one-dimensional compactified internal space. We find that the
energy scale for inflation 
can be predicted from the fine-structure constant
in a self-consistent solution of the semi-classical Einstein
equations including the Casimir effect. From
the observed value of the fine-structure constant, we
obtain an energy scale for inflation of $\epsilon=1.84\times
10^{16}g_*^{1/4}$~Gev, where $g_*$ is a dimensionless number depending 
on the spin and number of matter
fields existing in the universe. This value is consistent
with the values often discussed for inflation and grand unification.
The wave function for this model
predicts a high probability for forming such universes, independent of
the value of the cosmological constant. The tunneling probability
favors the creation of inflationary universes with a compactified
dimension, over those with all macroscopic dimensions.
\end{abstract}


Kaluza-Klein theory is a five-dimensional theory of gravity and 
electrodynamics, which is a combination of Einstein's
theory of gravity and Maxwell's theory of electrodynamics \cite{kal21}. 
In this theory the electric charge is quantized and the 
fine-structure constant $\alpha=e^2$ is determined by
the circumference $b$ of the 
one-dimensional internal space via $\alpha=64\pi^3 G/b^2$ where 
$G$ is the four-dimensional Newton's gravitational constant which is 
related to the five-dimensional gravitational constant $G_5$ 
via $G_5=Gb$.
(Throughout we use units where $c=\hbar=1$. If instead of the electron charge
$e$, considering quarks,  we use the electric charge $e/3$ as 
the fundamental charge, the
value of $b$ will be three times larger than the value we will quote.)
In this theory, the 
five-dimensional gravitational constant $G_5$ is the unique 
fundamental coupling constant, the four-dimensional gravitational constant
and the fine-structure constant are determined by the value of $G_5$
and the circumference of the internal space. To keep the four-dimensional
Newton's gravitational constant and the fine-structure constant
invariant with time, 
the circumference of the internal space must not change
with time. Modern accelerators have probed matter at scales as small as
$10^{-16}$~cm without finding any evidence for internal dimensions
\cite{kol90}. By contrast, our four-dimensional spacetime has a spatial scale 
as large as (or greater than) $10^{28}$~cm. If both the external
and internal (if it exists) scales originate from the Planck scale
($\sim10^{-33}$~cm) as modern cosmology suggests, the rate of change in 
the internal scale must be sufficiently small compared with the expansion
of the four-dimensional observed universe. Observations limiting the
variation of coupling constant with time also place strict restriction
on the rate of change in internal dimensions. Primordial nucleosynthesis
in Kaluza-Klein theory  implies that $1.01\ge b_{\rm N}
/b_0\ge0.99$ where $b_N$
is the circumference of internal space at the epoch of primordial
nucleosynthesis and $b_0$ is the circumference of the internal space today
\cite{kol90,kol86}. (By contrast, for the expand scale factor 
$a$ of the observed
four-dimensional universe, in the Big Bang theory we have $a_{\rm N}/a_0
\sim 10^{-10}$.) Thus, naturally, in cosmology
with extra dimensions people try to find solutions with the external
dimensions expanding while the internal dimensions remain static. But at
present no mechanism for keeping the internal spatial scale static has 
been found. In this paper, we give a model of an inflationary universe
in Kaluza-Klein theory, which is static in the one-dimensional internal
dimension while expanding in the external dimensions. We
find that this model can solve the semi-classical Einstein equations with the
Casimir effect \cite{cas48} or
vacuum polarization considered, and the fine-structure
constant $\alpha$ is related to the energy scale for inflation, $\epsilon$,
via $\epsilon/\epsilon_{\rm p}\simeq0.0176g_*^{1/4}\alpha^{1/2}$,
where $\epsilon_{\rm p}=G^{-1/2}$
is the Planck energy, and
$g_*$ is a dimensionless number determined by the spin 
and number of matter fields in the early universe ($g_*\sim100$). 
The idea that the energy-momentum
tensor required to produce the geometry
of spacetime with internal dimensions (via Einstein equations)
is provided by the Casimir effect 
plus a cosmological constant has been investigated by 
Weinberg \cite{wei83} and Candelas and Weinberg \cite{can84} 
for the case of four-dimensional flat 
Minkowski spacetime plus extra dimensions.
They have found that in order to
produce a reasonable value of the gauge coupling constant,
an enormous number (greater than 1000) of
matter fields are needed which is not supported by 
observations. Our model is a 
four-dimensional de Sitter space plus a compactified
one-dimensional flat internal space. 
For this model, we find that in order to produce the correct value of the 
fine-structure constant $\alpha=1/137.036$ it is 
required that the energy scale for inflation is $\epsilon=1.84\times10^{16}
g_*^{1/4}$~Gev which for any reasonable value of $g_*\sim 1-100$
is consistent with values often discussed for
inflation and grand unification \cite{bun96,kol90}. 

Our model is $dS^4\times S^1$ where $dS^4$ is a four-dimensional 
de~Sitter Space and $S^1$ is a compactified one-dimensional flat Euclidean 
space with circumference $b$. The metric is
\begin{eqnarray}
        ds^2=-d\tau^2+r_0^2\cosh^2 {\tau\over r_0}~\left[d\psi^2+
        \sin^2\psi~\left(d\theta^2+\sin^2\theta~d\phi^2\right)\right]
        +dq^2,
        \label{des}
\end{eqnarray}
where $(\tau,\psi,\theta,\phi)$ are the usual global coordinates on $dS^4$, 
$r_0$ is the radius of $dS^4$, $q$ is the Cartesian coordinate on $S^1$, 
and $(\tau,\psi,\theta,\phi,q)$ is identified with $(\tau,\psi,\theta,
\phi,q+nb)$ where $n=0,\pm1,\pm2,...$ Extending to the Euclidean section 
by setting $\tau\rightarrow-ir_0(\chi-\pi/2)$, $dS^4\times S^1$ is extended to
$S^4\times S^1$ where $S^4$ is a four-sphere with radius $r_0$ 
embedded in a five-dimensional flat Euclidean space. The corresponding 
metric extended from Eq.~(\ref{des}) is
\begin{eqnarray}
        ds^2=r_0^2\left\{d\chi^2+\sin^2 {\chi}~\left[d\psi^2+
        \sin^2\psi~\left(d\theta^2+\sin^2\theta~d\phi^2\right)\right]
        \right\}
        +dq^2,
        \label{edes}
\end{eqnarray}
where $(\chi,\psi,\theta,\phi)$ are spherical coordinates on $S^4$ with 
$0<\chi<\pi$, $0<\psi<\pi$, $0<\theta<\pi$, and $0\le\phi<2\pi$. Again, 
$(\chi,\psi,\theta,\phi,q)$ is identified with 
$(\chi,\psi,\theta,\phi,q+nb)$. A prominent feature of this model
is that, in the Lorentzian section $dS^4\times S^1$, the four-dimensional
external spacetime (the de~Sitter space) is exponentially expanding
while the one-dimensional compactified internal space $S^1$
is static; in the Euclidean section $S^4\times S^1$, both the external
space $S^4$ and the internal space $S^1$ are static and can have comparable
scales.

Various Lorentzian spacetimes can be obtained from different 
continuations of the Euclidean space $S^4\times S^1$:
1)~if we make the continuation $\chi\rightarrow \pi/2+i\tau/r_0$, 
$S^4$ is extended 
to $dS^4$ with the global coordinates having
closed spatial sections, which can 
be used to describe a closed universe created from nothing 
\cite{vil82,har83,lin84}; in our Kaluza Klein model,
$S^4\times S^1$ is extended to $dS^4\times S^1$, 
which is a four-dimensional closed de~Sitter universe with a compactified 
one-dimensional internal space and can be used to describe the creation 
of a closed five-dimensional inflationary universe with one dimension 
compactified and static; 2)~if 
we first make the continuation $\psi\rightarrow \pi/2+i\tau$ and then
make the continuation $\chi\rightarrow 
it$ and $\tau\rightarrow i\pi/2+\tilde\psi$ \cite{haw98} $S^4$ is extended 
to $dS^4$ with an open spatial section describing an open inflation 
\cite{got82,buc95,rat95,lin95}, which can be used to describe the creation 
of an open inflationary universe created from nothing \cite{haw98}; 
in this case, in our Kaluza-Klein model, $S^4\times S^1$ is extended 
to an open de~Sitter universe with a compactified one-dimensional internal 
space, i.e. creation of an open 
inflationary universe with a static compactified one-dimensional 
internal space (see \cite{gar98} for a non-static internal dimension
case); 3)~if we make the continuation $\phi\rightarrow it$, 
$S^4$ is extended to a de~Sitter space with closed timelike 
curves (CTCs) which 
describes a universe created from itself \cite{got98}; in this case in
our Kaluza-Klein model $S^4\times S^1$ is 
extended to a de~Sitter space with CTCs and a static and 
compactified one-dimensional 
internal space; 4)~if we make the continuation $q\rightarrow it$, 
$S^4\times S^1$ is extended to a five-dimensional Einstein static 
universe with CTCs (or without CTCs if $S^1$ is unfolded). Thus, 
we make our calculations in the Euclidean section $S^4\times S^1$ and 
then continue the results to the Lorentzian spacetimes we are interested in 
(except for the response function of a particle detector, which has to be 
calculated in the appropriate Lorentzian spacetime).

The space $S^4\times S^1$, with the metric in Eq.~(\ref{edes}), is 
conformally flat. With the coordinate transformation
\begin{eqnarray}
        r={\sin\chi\over2\left[\cos\chi+\cosh\left(q/r_0\right)\right]},~~~
        p={\sinh\left(q/r_0\right)\over2\left[\cos\chi+\cosh\left(
        q/r_0\right)\right]},
        \label{tra}
\end{eqnarray}
the metric in Eq.~(\ref{edes}) can be written as 
$ds^2=\Omega^2d\overline{s}^2$, 
where $\Omega=2r_0\left[\cos\chi+\cosh\left(q/r_0\right)\right]$ and 
$d\overline{s}^2$ is the metric of the five-dimensional flat Euclidean space
\begin{eqnarray}
        d\overline{s}^2=dr^2+r^2\left[d\psi^2+\sin^2\psi~\left(d\theta^2+
        \sin^2\theta~d\phi^2\right)\right]+dp^2.
        \label{emin}
\end{eqnarray}
Thus $S^4\times S^1$ is conformally flat.
Consider a massless and conformally coupled scalar field. The Hadamard 
function for the Minkowski vacuum in five-dimensional flat Euclidean space is
\begin{eqnarray}
        \overline{G}_{\rm M}^{(1)}\left(X,X^\prime\right)=
	{1\over4\pi^2}~{1\over\sigma^3}=
	{1\over4\pi^2}~{1\over
        \left[r^2+{r^\prime}^2-2rr^\prime\cos\Theta_3+\left(p-p^\prime\right)^2
        \right]^{3/2}},
        \label{gemin}
\end{eqnarray}
where $X=(r,\psi,\theta,\phi,p)$, $X^\prime=(r^\prime,
\psi^\prime,\theta^\prime,
\phi^\prime,p^\prime)$, $\sigma$ is the Euclidean distance between
$X$ and $X^\prime$, and $\cos\Theta_3=\cos\psi\cos\psi^\prime+\sin\psi
\sin\psi^\prime\cos\Theta_2$ with $\cos\Theta_2=\cos\theta\cos\theta^\prime+
\sin\theta\sin\theta^\prime\cos\left(\phi-\phi^\prime\right)$. 
The corresponding 
Hadamard function for the conformal Minkowski vacuum in $S^4\times R^1$ 
with the metric in Eq.~(\ref{edes}) is given by $G^{(1)}
\left(X,X^\prime\right)=
\Omega^{-3/2}\left(X\right)\overline{G}^{(1)}\left(X,X^\prime\right) 
\Omega^{-3/2}\left(X^\prime\right)$ \cite{bir82}, which is
\begin{eqnarray}
        G_{\rm CM}^{(1)}\left(X,X^\prime\right)=
        {1\over2^{7/2}\pi^2r_0^3}~{1\over
        \left\{\cosh\left[\left(q-q^\prime\right)/r_0\right]-\cos\Theta_4
        \right\}^{3/2}},
        \label{gedes}
\end{eqnarray}
where 
$X=(\chi,\psi,\theta,\phi,q)$, $X^\prime=(\chi^\prime,\psi^\prime,
\theta^\prime,\phi^\prime,q^\prime)$, and $\cos\Theta_4=\cos\chi\cos
\chi^\prime+\sin\chi\sin\chi^\prime\cos\Theta_3$ ($\Theta_4$ is
the angular distance between $X$ and $X^\prime$ on $S^4$).
The Hadamard function 
for the massless conformally coupled scalar field in the adapted 
conformal Minkowski vacuum (i.e. the conformal Minkowski vacuum with 
multiple images) in the Euclidean space $S^4\times S^1$ is given by the 
summation of Eq.~(\ref{gedes}) with multiple images
\begin{eqnarray}
        G^{(1)}\left(X,X^\prime\right)=
        {1\over2^{7/2}\pi^2r_0^3}~\sum_{n=-\infty}^{\infty}{1\over
        \left\{\cosh\left[\left(q-q^\prime+nb\right)/r_0\right]-\cos\Theta_4
        \right\}^{3/2}}.
        \label{gedes1}
\end{eqnarray}
The expectation value of the energy-momentum tensor is obtained from the 
Hadamard function by
\begin{eqnarray}
        \langle T_{ab}\rangle={1\over2}\lim_{X^\prime\rightarrow X}D_{ab}
        G^{(1)}(X,X^\prime),
        \label{tab}
\end{eqnarray}
where for a massless conformally coupled scalar field in a five-dimensional
spacetime
\begin{eqnarray}
        D_{ab}={5\over8}\nabla_{a^\prime}\nabla_b-{3\over8}\nabla_a
        \nabla_b+{3\over8}g_{ab}\left(g^{cd}\nabla_c\nabla_d-{1\over3}
        g^{cd}\nabla_{c^\prime}\nabla_d\right)+{3\over16}\left(R_{ab}-{1\over2}
        Rg_{ab}\right),
        \label{dab}
\end{eqnarray}
where $R_{ab}$ is the Ricci tensor and $R=g^{ab}R_{ab}$. Inserting 
Eq.~(\ref{gedes1}) into Eq.~(\ref{tab}), the $n=0$ term diverges as 
$X^\prime\rightarrow X$, thus renormalization must be taken. However, here 
the renormalization is simplified by noting that the $n=0$ term in 
Eq.~(\ref{gedes1}) is just the Hadamard function in $S^4\times R^1$ given 
in Eq.~(\ref{gedes}). For the massless conformally coupled scalar field 
in odd dimensional conformally related spaces, there is no trace anomaly and 
the renormalized energy-momentum tensors are related by \cite{dew79}
\begin{eqnarray}
        \langle T_a^{~b}\rangle_{\rm ren}=\Omega^{-4}
        \langle\overline{T}_a^{~b}\rangle_{\rm ren}.
        \label{ren}
\end{eqnarray}
For the Minkowski vacuum in $R^5$ with metric in Eq.~(\ref{emin}) we have 
$\langle\overline{T}_a^{~b}\rangle_{\rm ren}=0$. Thus for the conformal 
Minkowski vacuum in $S^4\times R^1$ with metric in Eq.~(\ref{edes}) we also 
have $\langle T_a^{~b}\rangle_{\rm ren}=0$, by Eq.~(\ref{ren}). The $n=0$ 
term's contribution to the renormalized energy-momentum tensor of the 
adapted conformal Minkowski vacuum in $S^4\times S^1$ happens 
to be the renormalized energy-momentum tensor in $S^4\times R^1$,
which is thus zero as discussed above. 
Therefore, all contributions to the renormalized energy-momentum tensor 
in $S^4\times S^1$ come from the $n\neq0$
terms in the expansion of the Hadamard function in Eq.~(\ref{gedes1}),
a purely Casimir effect. Inserting all $n\neq0$ terms 
in Eq.~(\ref{gedes1}) into Eq.~(\ref{tab}), we obtain the renormalized 
energy-momentum tensor of the adapted 
conformal Minkowski vacuum in $S^4\times S^1$
\begin{eqnarray}
        \langle T_{\mu}^{~\nu}\rangle_{\rm ren}={3A\over8\pi r_0^5}
        \left(\begin{array}{ccccc}
        1&0&0&0&0\\
        0&1&0&0&0\\
        0&0&1&0&0\\
        0&0&0&1&0\\
        0&0&0&0&-4
        \end{array}
	\right),
        \label{tmn}
\end{eqnarray}
where $A=
{1\over6\pi}\sum_{n=1}^\infty{3\sinh^2{n\beta}+4\over\sinh^5{n\beta}}$ 
and $\beta=b/(2r_0)$, the coordinates are $(\chi,\psi,\theta,\phi,q)$.
The renormalized energy-momentum tensor in Eq.~(\ref{tmn})
has the form of radiation if we regard $q$ as Euclidean time.

The Ricci tensor of $S^4\times S^1$ with metric in Eq.~(\ref{edes}) is
$R_{ab}=\left(3/r_0^2\right)\left(g_{ab}-dq_adq_b\right)$.
The Ricci scalar is $R=g^{ab}R_{ab}=12/r_0^2$. The five-dimensional 
semi-classical Einstein equations are
\begin{eqnarray}
        R_{ab}-{1\over2}Rg_{ab}+\Lambda g_{ab}=8\pi G_5\langle
        T_{ab}\rangle_{\rm ren},
        \label{ein}
\end{eqnarray}
where $\Lambda$ is the cosmological constant and $G_5$ is the 
five-dimensional gravitational constant. Inserting Eq.~(\ref{tmn})  
into Eq.~(\ref{ein}), we find that the semi-classical 
Einstein equations are solved if
\begin{eqnarray}
        r_0^3=5G_5A,
        \label{rrr}
\end{eqnarray}
and  $\Lambda=18G_5A/r_0^5=18/5r_0^2$.
The asymptotic solution of Eq.~(\ref{rrr}) as $r_0\rightarrow\infty$ is 
$b=\left(10G_5/\pi\right)^{1/5}r_0^{2/5}$. Numerical
calculation shows that this solves Eq.~(\ref{rrr}) very accurately for any 
$r_0>{\overline{l}_{\rm p}}$ where $\overline{l}_{\rm p}=G_5^{1/3}$ 
is the five-dimensional Planck length. For example, for 
$r_0={\overline{l}_{\rm p}}$, the asymptotic solution gives $b=1.2606
~{\overline{l}}_{\rm p}$, while the numerical solution is $b=1.2594
~{\overline{l}}_{\rm p}$, the relative error is only $0.1\%$. 
Thus the asymptotic 
solution solves Eq.~(\ref{rrr}) very accurately for $r_0>
{\overline{l}_{\rm p}}$. 
If there are many conformally coupled matter fields instead of only one
scalar field, the renormalized energy-momentum tensor is 
given by Eq.~(\ref{tmn})
multiplied by a factor $g_*$ determined by the 
number of species and spins of the
matter fields. Then the asymptotic solution is
\begin{eqnarray}
	{b\over\overline{l}_{\rm p}}=\left({10g_*\over\pi}\right)^{1/5}
	\left({r_0\over\overline{l}_{\rm p}}\right)^{2/5}
        \simeq1.26g_*^{1/5}
	\left({r_0\over\overline{l}_{\rm p}}\right)^{2/5}.
	\label{bbb}
\end{eqnarray}
Using the four-dimensional Planck length $l_{\rm p}=G^{1/2}$ and 
$G=G_5/b$, Eq.~(\ref{bbb}) can also be written as
\begin{eqnarray}
	{b\over l_{\rm p}}=\left(10g_*\over\pi\right)^{1/4}
	\left({r_0\over l_{\rm p}}\right)^{1/2}\simeq
	1.336g_*^{1/4}\left({r_0\over l_{\rm p}}\right)^{1/2}.
	\label{bb1}
\end{eqnarray}
  
The fine-structure constant is $\alpha=64\pi^3G/b^2=64\pi^3\left(l_{\rm p}/b
\right)^2$. For an observer in $dS^4\times S^1$ 
which is a Lorentzian extension 
of $S^4\times S^1$, the effective cosmological constant is $\Lambda_{\rm eff}=
3/r_0^2=(5/6)\Lambda$. Thus Eq.~(\ref{bb1}) gives a correlation 
between $\Lambda_{\rm eff}$, $\alpha$ and $G$
\begin{eqnarray}
	G\Lambda_{\rm eff}={15g_*\over2048\pi^7}\alpha^2.
	\label{lam1}
\end{eqnarray}
In inflation theory 
people usually use an inflation potential instead 
of the effective cosmological
constant. The inflation potential is related to 
the effective cosmological
constant via $\Lambda_{\rm eff}=8\pi GV$. 
If we write the inflation potential
as $V=\epsilon^4$ where $\epsilon$ is the 
energy scale for inflation, 
Eq.~(\ref{lam1}) thus gives a correlation between the energy 
scale of inflation and the fine-structure constant
\begin{eqnarray}
	{\epsilon\over\epsilon_{\rm p}}={15^{1/4}g_*^{1/4}\over2^{7/2}\pi^2}
	\alpha^{1/2}\simeq0.0176g_*^{1/4}\alpha^{1/2},
        \label{eee}
\end{eqnarray}
where $\epsilon_{\rm p}=G^{-1/2}$ is the Planck energy. 
If at the epoch of inflation the fine-structure constant 
and the four-dimensional 
gravitational constant have the same values as today 
(as argued in the beginning
of the paper), we have $\epsilon\simeq1.84\times10^{16}g_*^{1/4}$~Gev and
$b=521l_{\rm p}=8.4\times 10^{-31}$~cm
(or $\epsilon\simeq0.61\times10^{16}g_*^{1/4}$~Gev and 
$b=1.56\times10^3l_{\rm p}
=2.5\times 10^{-30}$~cm if charge is quantized in units of $e/3$). 
The value of $\epsilon$ is 
not very sensitive to $g_*$ which is expected to be of order $10^2$
and is an energy scale often discussed for inflation and grand
unification \cite{bun96,kol90}. 

$S^4\times S^1$ is a compact five-dimensional Euclidean space. The
Euclidean action of a compact manifold plays an important role in
Hartle and Hawking's no-boundary proposal for quantum cosmology, which 
gives the probability for a universe created from nothing \cite{har83}.
With the Wick rotation $t\rightarrow-i\tau$, the Euclidean Hilbert-Einstein
action is $I_{\rm g}=-\left(1/16\pi G_5\right)\int d^5x\sqrt{g}
\left(R-2\Lambda
\right)$. For $S^4\times S^1$, we have $R=12/r_0^2$ and thus
$I_{\rm g}=-\left(1/16\pi G_5\right)\left(12/r_0^2-2\Lambda\right)V_5$
where $V_5=8\pi^2r_0^4b/3$ is the volume of $S^4\times S^1$. For the
self-consistent (i.e. the Einstein equations are solved) case, we
have $\Lambda=18/5r_0^2$, thus $I_{\rm g}=-4\pi r_0^2b/5G_5$. 
The vacuum polarization of conformal
fields (which is pure Casimir effect for the case 
of $S^4\times S^1$) gives rise in the Euclidean regime
to a radiation-like energy-momentum tensor [Eq.~(\ref{tmn})], whose Euclidean
action is $I_{\rm m}=\int d^5x\sqrt{g}\rho$ where 
$\rho$ is the ``energy" density
(in this case it is minus the $q-q$ component of $\langle T_{\mu}^{~\nu}
\rangle_{\rm ren}$). From Eq.~(\ref{tmn}) we have $\rho=3A/2\pi r_0^5 =
3/\left(10\pi G_5r_0^2\right)$ where Eq.~(\ref{rrr}) has been used, 
thus $I_{\rm m}=\rho V_5=4\pi r_0^2b/5G_5$. 
Notice that $I_{\rm m}$ and $I_{\rm g}$
have the same magnitude but opposite sign, thus the total
Euclidean action is $I=I_{\rm g}+I_{\rm m}=0$ (this result is 
independent of $g_*$), and thus the probability for
the creation of the universe from nothing is $P=1$ in the
Hartle and Hawking formulation. This means that the creation
of universes with different radius $r_0$ (and thus different effective
cosmological constant $\Lambda_{\rm eff}=3/r_0^2$) has equal probability.

For a 
five-sphere $S^5$ with metric $ds^2=r_0^2\left(d\alpha^2+\sin^2\alpha~
d\Theta_4^2\right)$ where $r_0$ is the radius of $S^5$, we have $R=20/r_0^2$
and $V_5=\pi^3r_0^5$. Such an $S^5$ is a Euclidean solution of the
five-dimensional vacuum Einstein 
equations with a cosmological constant $\Lambda=6/r_0^2$. (The renormalized
energy-momentum tensor is zero for a conformally coupled scalar field
in the conformal Minkowski vacuum in $S^5$ due to the fact
$S^5$ is conformally flat and there is
no trace anomaly in odd-dimensional spaces.) Thus for $S^5$ we have 
$I=I_{\rm g}=-\left(1/16\pi G_5\right)\int d^5x\sqrt{g}(R-2\Lambda)
=-\left(\pi^2/2G_5\right)r_0^3=-\left(\pi^2/2G_5\right)
\left(6/\Lambda\right)^{3/2}$. The probability for the creation of a universe
from an instanton is given by $P=e^{-|I|}$. (For the necessity of the absolute 
value of $I$ in the expression for $P$, see \cite{lin84,far90,lin98}. 
This is obviously true for quantum tunneling processes in ordinary 
quantum mechanics in flat spacetime.) 
Thus the probability for the creation of a universe
from $S^5$ is $P=\exp\left[-\left(\pi^2/2\right)\left(r_0/\overline{l}_{\rm p}
\right)^3\right]=\exp\left[-\left(\pi^2/2G_5\right)\left(6/
\Lambda\right)^{3/2}\right]$, which is always less than unity (for $r_0>
\overline{l}_{\rm p}$ we have $P<0.7\%$). Therefore the creation of a
universe from $S^4\times S^1$ (which has a probability equal to unity) is
more probable than from $S^5$, which could explain why we are not living
in a universe with five macroscopic dimensions.

Suppose there is a point-like particle detector moving along a geodesic 
in the spacetime of
$dS^4\times S^1$, with $q={\rm constant}$. Since the vacuum state is 
de~Sitter invariant, we can always choose global
coordinates in de~Sitter space such that $\psi={\rm constant}$, $\theta={\rm
constant}$, and $\phi={\rm constant}$ along the 
geodesic. The response function of the 
particle detector is \cite{bir82}
\begin{eqnarray}
   {\cal F}(\Delta E)=\int_{-\infty}^{\infty}d\tau\int_{-\infty}^\infty
   d\tau^\prime
   e^{-i\Delta E(\tau-\tau^\prime)}G^+(X(\tau),X(\tau^\prime)),
   \label{res}
\end{eqnarray}
where $G^+$ is the Wightman function of the detected field, $\tau$ is the 
proper time of the detector's
worldline, $\Delta E$ is the energy difference between an excited state of the 
detector and its ground state. The 
Hadamard function in $dS^4\times S^1$, in the
global de~Sitter coordinates,  is obtained
from Eq.~(\ref{gedes1}) by extending $\chi\rightarrow\pi/2+i\tau/r_0$. The
Wightman function is equal to one 
half of the corresponding Hadamard function with
$\tau$ replaced by $\tau-i\varepsilon/2$ and 
$\tau^\prime$ replaced by $\tau^\prime+
i\varepsilon/2$ where $\varepsilon$ is a positive infinitesimal real number
\cite{li98,got98}. On the worldline of the particle detector,
the Wightman function so obtained is
\begin{eqnarray}
	G^+(\Delta\tau)={1\over2^{9/2}\pi^2r_0^3}\sum_{n=-\infty}^\infty{1\over
	\left\{\cosh\left(nb/r_0\right)-\cosh\left[\left(
	\Delta\tau-i\varepsilon\right)
	/r_0\right]\right\}^{3/2}},
	\label{wigh}
\end{eqnarray}
where $\Delta\tau=\tau-\tau^\prime$. 
Inserting Eq.~(\ref{wigh}) into Eq.~(\ref{res}),
the integral of the $n=0$ term can be worked 
out with the residue theorem by choosing
a contour closed in the lower-half plane of 
complex $\Delta\tau$. The result is a 
Fermi-Dirac-like distribution with temperature 
$T=1/2\pi r_0$ (the Boltzmann constant is taken to be unity)
\begin{eqnarray}
	{d{\cal F}\over d\overline{\tau}}={1\over8\pi r_0^2}~
	{\left(\Delta Er_0\right)^2+1/4\over e^{2\pi\Delta Er_0}+1},
	\label{f00}
\end{eqnarray}
where $\overline{\tau}=\left(\tau+\tau^\prime
\right)/2$. (Similar conclusions for
an accelerated particle detector in odd dimensional Minkowski spaces can be
found in \cite{tak85}.) The terms with $n\neq0$ cannot
be worked out with the residue theorem because a closed 
contour on which the integrand 
is analytic does not exist. The integrals of the
$n\neq0$ terms can only be worked out
numerically. However, it can be estimated that the 
contributions of the $n\neq0$
terms are negligible compared to the $n=0$ term's 
contribution, because, for large $\Delta
Er_0$, the oscillation of the integrand 
causes the contributions from different terms
with different $n$ to cancel each other; 
for small $\Delta Er_0$, the contribution
of the $n\neq0$ terms is of the order of 
$\left(\Delta Er_0\right)^2$. Thus the $n=0$
term dominates the contribution to the response function. 

Thus, consideration of Kaluza-Klein theory in terms of inflation offers some
promising features. Given the observed value of the fine-structure constant
a self-consistent solution including the Casimir effect for an early 
inflationary state suggests an effective value for the cosmological constant
which is quite reasonable (corresponding to an energy scale of $1.84\times
10^{16}g_*^{1/4}$~Gev where $g_*\sim100$). Interestingly, when the
renormalized energy-momentum tensor due to the Casimir effect is included
in the Euclidean action it makes tunneling to create an inflationary universe
with a compactified dimension quite probable.

~

This work is supported by NSF grant AST95-29120 and NASA grant NAG5-2759.

\end{document}